\documentclass[twocolumn,superscriptaddress,amsmath,amssymb]{revtex4}

\usepackage{graphicx}
\usepackage{dcolumn}
\usepackage{bm}
\usepackage{ulem}
\newcommand{\nc}{\newcommand}
\nc{\be}{\begin{eqnarray}}
\nc{\ee}{\end{eqnarray}}
\nc{\bea}{\begin{eqnarray}}
\nc{\eea}{\end{eqnarray}}
\nc{\bean}{\begin{eqnarray*}}
\nc{\eean}{\end{eqnarray*}}

\begin{document}

\title{Evidence of a universal relation between electron-mode coupling and $T_c$ in Ba$_{1-x}$K$_x$Fe$_2$As$_2$ superconductor from Laser ARPES}




\author{W.~Malaeb}
\affiliation{Institute for Solid State Physics (ISSP), University of
Tokyo, Kashiwa-no-ha, Kashiwa, Chiba 277-8561, Japan}

\author{T.~Shimojima}
\affiliation{Department of Applied Physics, University of Tokyo, Tokyo 113-8656, Japan}
\affiliation{CREST, JST, Chiyoda-ku, Tokyo 102-0075, Japan}

\author{Y.~Ishida}
\affiliation{Institute for Solid State Physics (ISSP), University of
Tokyo, Kashiwa-no-ha, Kashiwa, Chiba 277-8561, Japan}
\affiliation{CREST, JST, Chiyoda-ku, Tokyo 102-0075, Japan}

\author{T.~Kondo}
\affiliation{Institute for Solid State Physics (ISSP), University of
Tokyo, Kashiwa-no-ha, Kashiwa, Chiba 277-8561, Japan}

\author{K.~Okazaki}
\affiliation{CREST, JST, Chiyoda-ku, Tokyo 102-0075, Japan}
\affiliation{Department of Physics, University of Tokyo, Tokyo 113-8656, Japan}

\author{Y.~Ota}
\affiliation{Institute for Solid State Physics (ISSP), University of
Tokyo, Kashiwa-no-ha, Kashiwa, Chiba 277-8561, Japan}

\author{K.~Ohgushi}
\affiliation{Institute for Solid State Physics (ISSP), University of
Tokyo, Kashiwa-no-ha, Kashiwa, Chiba 277-8561, Japan}

\author{K.~Kihou}
\affiliation{National Institute of Advanced Industrial Science and
Technology (AIST), Tsukuba, Ibaraki 305-8568, Japan}

\author{C.H.~Lee}
\affiliation{National Institute of Advanced Industrial Science and
Technology (AIST), Tsukuba, Ibaraki 305-8568, Japan}

\author{A.~Iyo}
\affiliation{National Institute of Advanced Industrial Science and
Technology (AIST), Tsukuba, Ibaraki 305-8568, Japan}

\author{H.~Eisaki}
\affiliation{National Institute of Advanced Industrial Science and
Technology (AIST), Tsukuba, Ibaraki 305-8568, Japan}

\author{S.~Ishida}
\affiliation{National Institute of Advanced Industrial Science and
Technology (AIST), Tsukuba, Ibaraki 305-8568, Japan}

\author{M.~Nakajima}
\affiliation{Department of Physics, Osaka University, Osaka 560-0043, Japan}

\author{S.~Uchida}
\affiliation{Department of Physics, University of Tokyo, Tokyo
113-8656, Japan}

\author{H.~Fukazawa}
\affiliation{Department of Physics, Chiba University, Chiba
263-8522, Japan}

\author{T.~Saito}
\affiliation{Department of Physics, Chiba University, Chiba
263-8522, Japan}

\author{Y.~Kohori}
\affiliation{Department of Physics, Chiba University, Chiba
263-8522, Japan}

\author{S.~Shin}
\affiliation{Institute for Solid State Physics (ISSP), University of
Tokyo, Kashiwa-no-ha, Kashiwa, Chiba 277-8561, Japan}
\affiliation{CREST, JST, Chiyoda-ku, Tokyo 102-0075, Japan}

\date{\today}



\begin{abstract}

We performed a Laser angle-resolved photoemission spectroscopy (ARPES) study on a wide doping range of Ba$_{1-x}$K$_x$Fe$_2$As$_2$ (BaK) iron-based superconductor. We observed a robust low-binding energy (BE) kink structure in the dispersion which is doping dependent where its energy peaks at the optimally-doped (OP) level ($x \sim$ 0.4) and decreases towards the underdoped (UD) and overdoped (OD) sides. It is also temperature-dependent and survives up to $\sim$ 90K. We attribute this kink to electron-mode coupling in good agreement with the inelastic neutron scattering (INS) and scanning tunneling microscopy (STM) results on the same compound which observed a similar bosonic mode associated with spin excitations. The relation between the mode energy ($\Omega$) and the SC transition temperature ($T_c$) deduced from our Laser ARPES data follow the universal relation deduced from INS and STM. In addition, we could resolve another kink at higher BE showing less doping and temperature dependence and may thus be of different origin.

\end{abstract}

\maketitle




There is still no full understanding of the origin of superconducting (SC) pairing mechanism in iron-superconductors (FeSCs) \cite{Kamihara_JACS08}. One dominant theoretical opinion suggests that the electron pairing is of magnetic origin induced by antiferromagnetic (AFM) spin-fluctuations (SF) between disconnected Fermi surface (FS) sheets \cite{Mazin_PRL08, Kuroki_PRL08, Kuroki_PRB09}. Inelastic neutron scattering (INS) data of several FeSCs have shown a resonance peak-like feature below the SC transition temperature ($T_c$) localized in both energy and wavevector \cite{Christianson_Nat08, Castellani_PRL11} which was attributed to a spin excitation mode. Moreover, recently a peak-dip-hump structure was observed in the STM data of several FeSCs \cite{Hoffman_Rev, HHWen_NatPhys13} which was taken as an evidence for a bosonic mode in these compounds identical to the spin excitation mode observed by INS experiments. This mode is believed to be in close relation to superconductivity and associated with spin excitations instead of phonons \cite{Hoffman_Rev, HHWen_NatPhys13}. Therefore, by now at least two different experimental techniques have shown evidence for bosonic mode coupling in FeSCs and demonstrated a universal relation between the mode energy ($\Omega$) and $T_c$ \cite{Hoffman_Rev, HHWen_NatPhys13}. In principle, this mode coupling should be directly observed by angle-resolved photoemission spectroscopy (ARPES) as an anomaly in the band dispersion (kink) where, compared to other experimental techniques, ARPES is considered as the most direct in revealing such features as has been clearly demonstrated in cuprates \cite{Lanzara, Damascelli, Terashima}.

Indeed, FeSCs have been extensively studied by ARPES especially Ba$_{1-x}$K$_x$Fe$_2$As$_2$ (BaK) compound \cite{Malaeb, Okazaki_Science12, Shimojima_Science11, Ding3D, Feng_PRL10, Borisenko_arXiv12, Borisenko_arXiv11, Nakayama, Richard, Ding} and the coupling to a bosonic mode has been already discussed and compared with INS data \cite{Richard, Borisenko_arXiv11}. However, most of these studies were mainly focused on the OP compound and thus no full information is available about the doping evolution of the bosonic mode in BaK similar to the information given by INS \cite{Castellani_PRL11} for example. Moreover, most of the previous ARPES data on BaK have shown a complicated line shape, specifically a two-peak feature in the energy distribution curves (EDCs) which was sometimes poorly resolved and appeared as a shoulder depending on the sample quality, photon energy, energy resolution and possibly other experimental conditions as well \cite{Malaeb, Okazaki_Science12, Shimojima_Science11, Ding3D, Feng_PRL10, Borisenko_arXiv12, Borisenko_arXiv11, Nakayama, Richard, Ding} causing complications in specifying the different energy scales in this compound. Therefore, we have employed Laser ARPES with photon energy $h\nu$= 6.994 eV and energy resolution of $\sim$3 meV to study a wide doping range of BaK from the underdoped (UD) to the overdoped (OD) regions to carefully investigate the different energy scales and the possibility of electron-boson coupling and its doping evolution in this compound.

Our ARPES data show a robust low-BE kink structure in the dispersion which is doping dependent with its energy peaking at the OP level ($x \sim$ 0.4) and decreasing towards the UD and OD sides. It is also temperature-dependent and survives up to $\sim$ 90K. We attribute this kink to electron-mode coupling in good agreement with INS and STM data on the same compound which observed a bosonic mode at a similar energy scale and was considered to be related with spin excitations. The relation between the mode energy ($\Omega$) and the SC transition temperature ($T_c$) deduced from our Laser ARPES data follow the universal relation deduced from INS and STM. In addition, we could resolve another kink at higher BE showing less doping and temperature dependence and may thus be of different origin. Such a clear separation of two kinks in BaK ARPES data attributed with different origins has not been clearly presented before. Instead, previous ARPES results \cite{Richard, Borisenko_arXiv11} discussed one kink in OP BaK with no clear information about its doping and temperature dependence.





High-quality single crystals of Ba$_{1-x}$K$_x$Fe$_2$As$_2$ (BaK) were grown by using the FeAs flux
\cite{Ohgushi} for the underdoped (UD) ($x \sim$ 0.3) and optimally-doped (OP) ($x \sim$ 0.4)
samples while the KAs \cite{Kihou} flux was used for overdoped (OD) ($x \sim$ 0.51, 0.57, 0.6 and 0.7) samples. Laser ARPES
measurements were carried out at ISSP, University of Tokyo using
a VG-Scienta R4000 as an electron analyzer and a VUV Laser of $h\nu$
= 6.994 eV as a light source. In this spectrometer \cite{Kiss},
changing the polarization vector of the laser source is possible by
rotating the half-wave ($\lambda$/2) plate without changing the
optical path. The Fermi level ($E_F$) calibration of the samples was
done by referring to that of gold and the energy resolution was set
to $\sim$3 meV. The samples were cleaved at a temperature $T \sim$
160 K or less in an ultra-high vacuum of less than 5 $\times$
10$^{-11}$ Torr.









\begin{figure}[htb]
\begin{center}
\includegraphics[width=9.5cm]{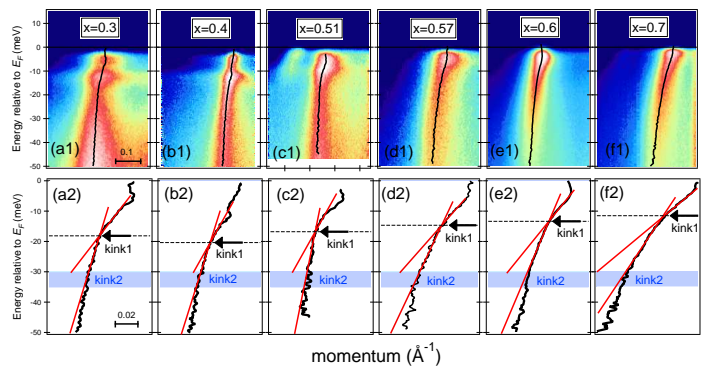}
\caption{\label{fig1} (color online) Laser ARPES data of BaK with different dopings from UD to OD: (a1-f1) $E-k$ plots of a cut taken near the high-symmetry line with v-polarized light at  $T\sim$ 7 K in the SC state of several doping levels. The black solid curves represent MDC peak positions determined from Lorentzian fitting which is displayed again in larger scale in panels (a2-f2) respectively. The low-BE kink1 energy position, marked by a black arrow, was estimated from the intersection points between two different slopes in the dispersion represented by red solid lines. The thick blue horizontal lines represent an additional high-BE kink2 in the dispersion.}
\end{center}
\end{figure}

Fig. \ref{fig1} displays the Laser ARPES data of BaK with several doping levels ranging from the underdoped (UD) to the overdoped (OD) side. $E$-$k$ plots corresponding to a cut taken with v-polarized light along or close to the high-symmetry line near the brillouin zone (BZ) center in the SC state at $T \sim$ 7 K are displayed in panels (a1-f1) corresponding to doping levels ranging from $x \sim$ 0.3 to $x \sim$ 0.7 respectively. Due to symmetry considerations and polarization direction of the incident light, these $E$-$k$ plots mainly display the inner hole-like band around the BZ center \cite{Malaeb, Shimojima_Science11, Okazaki_Science12}. Whereas a simple Bogoliubov quasiparticle dispersion is observed in the OD region similar to the case of cuprates \cite{Lanzara, Damascelli, Terashima}, the dispersion becomes complicated in the UD and OP regions, however the analysis which seems straightforward in the OD region is also valid in the UD and OP regions. Now we overlap on each of the $E$-$k$ plots of Fig. \ref{fig1} the peak positions of momentum distribution curves (MDCs) determined from Lorentzian fitting represented by black solid curves. In order to take a closer look and check for possible anomalies in the dispersion, we display in panels (a2-f2) the same curves but with an enlarged scale. We check the slope of the black solid curves representing the dispersion for each doping and we find out that it is a non-monotonic slope which can be seen more clearly by observing the red solid lines overlapped on the two different slopes close to the Fermi level ($E_F$) where we focus our discussion. The change in the slope of the dispersion implies the presence of an anomaly or a kink which we label as kink1 and mark its position by a black arrow. The intersection points between the two red solid lines representing the different slopes in the dispersion is the kink energy ($E_{kink}$). Such analysis method is frequently used especially in the case of cuprates \cite{Lanzara, Damascelli, Terashima}. Now, given that many-body interactions like the electron-mode coupling usually appear as anomalies in the photoemission spectrum, here we consider the anomaly in the dispersions (kink1) of Fig. \ref{fig1} to be originating from such electron-mode coupling. Before discussing the doping- and temperature-dependence of this mode as well as its possible origin, we note that we could resolve an additional kink at higher BE ($ \sim$ 30 - 35 meV) which we label as kink2 and mark by a blue thick horizontal line. Compared to kink1, this additional kink2 does not show much doping-dependence as can be seen in Fig. \ref{fig1} or temperature dependence as we shall see later in Fig. \ref{fig3}. Such a clear separation of two kinks in BaK ARPES data has not been clearly shown before. Instead, previous ARPES results \cite{Richard, Borisenko_arXiv11} discussed one kink in OP BaK with no clear information about its doping and temperature dependence. From this point ahead our discussion will be mainly focused on kink1.



\begin{figure}[htb]
\begin{center}
\includegraphics[width=9.5cm]{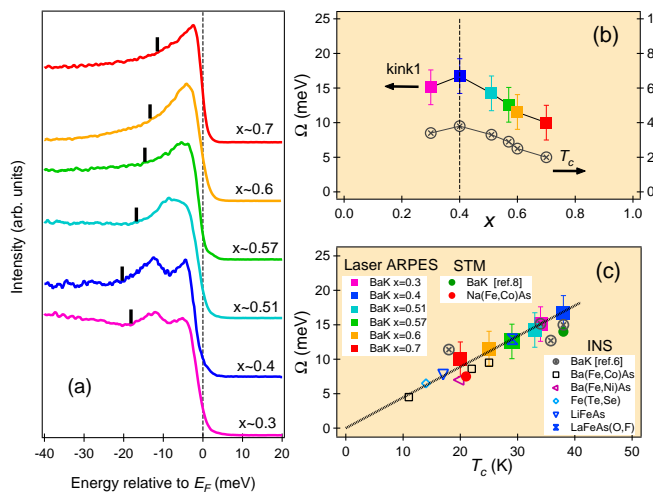}
\caption{\label{fig2} (color online) (a) Inner hole-like band EDCs in the SC state of BaK with several doping levels ranging from the UD to the OD region corresponding to the $E-k$ plots in Fig. \ref{fig1}. The kink energy $E_{kink}$ corresponding to the mode and which was determined from Fig. \ref{fig1} is marked by black bars. (b) Doping dependence of the mode energy $\Omega$ (left $y$ axis) and the SC transition temperature $T_c$ (right $y$ axis). (c) The mode energy $\Omega$ versus $T_c$ in BaK. The black dotted line is a linear fit of the data points implying the ratio $\Omega$/$k_B$$T_c$ $\sim$ 5.16. Overlapped on our Laser ARPES points are STM and INS data points corresponding to several FeSC compounds including BaK taken from literature \cite{Castellani_PRL11, Hoffman_Rev, HHWen_NatPhys13}. Notice the color correspondence in all three panels of this figure.}
\end{center}
\end{figure}




The doping dependence of the low-BE kink structure (kink1) and the corresponding mode are displayed in Fig. \ref{fig2}. Panel (a) shows the EDCs at $k_{F}$ of the inner hole band in the SC state corresponding to the $E$-$k$ plots of Fig. \ref{fig1}. Using the analysis in Fig. \ref{fig1} (a2-f2), we determined the kink energy ($E_{kink}$) as the intersection point between the two red solid lines for each doping and this energy position $E_{kink}$ is marked by a black bar on each of the EDCs of Fig. \ref{fig2}(a). Now, considering that we have assigned kink1 as an effect of electron-mode coupling we proceed to determine the mode energy ($\Omega$) for each doping level. Following a similar analysis to the case of cuprates \cite{Lanzara, Damascelli, Terashima}, we determined $\Omega$ by subtracting the SC gap size ($\Delta$) from the kink energy ($E_{kink}$) value for each EDC ($\Omega$=$E_{kink}$-$\Delta$). We note that the SC gap size was estimated by fitting the low-BE peak in all the EDCs in Fig. \ref{fig2} (a) by a BCS spectral function similar to Ref. \cite{Malaeb, Shimojima_Science11}. We then plotted the obtained $\Omega$ values versus $x$ in Fig. \ref{fig2}(b): a clear doping dependence of the mode energy $\Omega$ is observed, where it peaks at the OP level ($x \sim$ 0.4) and decreases toward the UD and OD sides in a similar fashion like $T_c$ which is also plotted in panel (b). We compare these results with the doping-dependence of the mode deduced from INS results on the same compound BaK \cite{Castellani_PRL11}: A good agreement is observed where the same energy scale of the mode with a similar doping dependence is observed in both ARPES and INS which strongly suggests that both could be detecting the same bosonic mode. Moreover, a similar mode was observed in OP BaK by STM \cite{HHWen_NatPhys13} as we mentioned earlier. In order to make a more clear comparison between all these experimental data, we first refer to the agreement between INS and STM data for BaK compound and other FeSCs that was already discussed and manifested in the universal relation between $\Omega$ and $T_c$ \cite{Hoffman_Rev, HHWen_NatPhys13} which is also valid for cuprates \cite{Hoffman_Rev}. We have checked to what extent our results agree with this universal relation by making a similar plot between $\Omega$ and $T_c$ in Fig. \ref{fig3}(c) where the different $T_c$'s in this case correspond to different doping levels of BaK rather than different compounds as was plotted in Ref.\cite{Hoffman_Rev}. To our knowledge, this is the first time such a plot is shown for BaK from ARPES data. It is remarkable that our plot follows a linear behavior and gives a ratio $\Omega$/$k_B$$T_c$ $\sim$ 5.16 comparable with the values obtained from INS and STM data on FeSCs including BaK ($\Omega$/$k_B$$T_c$ $\sim$ 4.4-4.8) \cite{Hoffman_Rev, HHWen_NatPhys13}. To show this more clearly, we overlap on the same graph data points taken from INS \cite{Castellani_PRL11} and STM \cite{HHWen_NatPhys13} results corresponding to several FeSCs including BaK. Interestingly, most of these data points fall on the same line like our Laser ARPES data points. This also reconfirms the unconventional superconductivity in FeSCs similar to the case of cuprates which also follow this universal relation between $\Omega$ and $T_c$ \cite{Hoffman_Rev}. All this gives a strong evidence that the mode observed in our Laser ARPES data could be the same as that observed in both INS and STM data which considered it to be associated with spin excitations. This shall be further clarified after presenting our temperature-dependent data in the next section.


But before, we note that an additional high-BE peak is observed in the EDCs of Fig. \ref{fig2}(a) especially near OP BaK located at $\sim$ 10-13 meV. This peak corresponds to the additional intensity observed in the $E$-$k$ plots of Fig. \ref{fig1} near OP doping and complicates the dispersion. Obviously, the intensity of this peak is more enhanced in the OP and UD regions compared with the OD region where it gradually diminishes and disappears at $x$ $\sim$ 0.6-0.7. Actually this two-peak feature is not only unique to Laser data but has been observed by previous ARPES studies using both He-discharge lamp and synchrotron radiation especially near OP BaK \cite{Ding3D, Feng_PRL10, Borisenko_arXiv12, Borisenko_arXiv11, Nakayama, Richard, Ding} and in fact it complicated the interpretation of the photoemission spectra in this compound. Indeed, the presence of such additional peak in BaK spectra is very unusual as compared to other FeSCs or to cuprates for example.






After presenting our doping-dependent Laser ARPES data and their good agreement with INS and STM results, we proceed to show the temperature-dependence of the kink structure in Fig. \ref{fig3} for $x \sim$ 0.4 sample. $E$-$k$ plots of a cut taken near the high-symmetry line with v-polarized light at different temperatures are presented in the upper panel. The data were divided by Fermi-Dirac (FD) function. The corresponding MDC peak positions determined from Lorentzian fitting (as in Fig. \ref{fig1}) are displayed as colored curves in the bottom panel of Fig. \ref{fig3}. From these curves, we determined the temperature-dependence of the low-BE kink1 which is the key for identifying the kink energy $E_{kink}$ from which we determined the mode energy $\Omega$. The kink energy, marked by a black arrow, was determined in a similar manner as in Fig. \ref{fig1}. From this plot, we deduce that kink1 survives up to $\sim$ 90K much higher than $T_c$ ($\sim$ 38K) and with the closure of the SC gap at $T_c$, its energy $E_{kink}$ is shifted towards $E_F$ by an amount equivalent to the SC gap $\Delta$ before it disappears at $\sim$ 90K. On the other hand, the high-BE kink2 which we mark by a thick blue horizontal line, shows weak temperature dependence and can still be observed much above $T_c$ (up to 170K). The different temperature dependencies of kinks (1) and (2) suggest that these two kinks may be of different origins. While we have attributed kink1 to a spin excitation mode, we suggest that kink2 may correspond to the phonon mode observed in the inelastic x-ray scattering data at $\sim$ 25-30 meV \cite{Lee} but this should be further investigated in future studies. We remark that, owing to the high-energy resolution of Laser ARPES we could discriminate the two kinks (1) and (2) and their different temperature dependencies which could not be achieved by previous ARPES studies on BaK \cite{Richard}.



\begin{figure}[htb]
\begin{center}
\includegraphics[width=9.3cm]{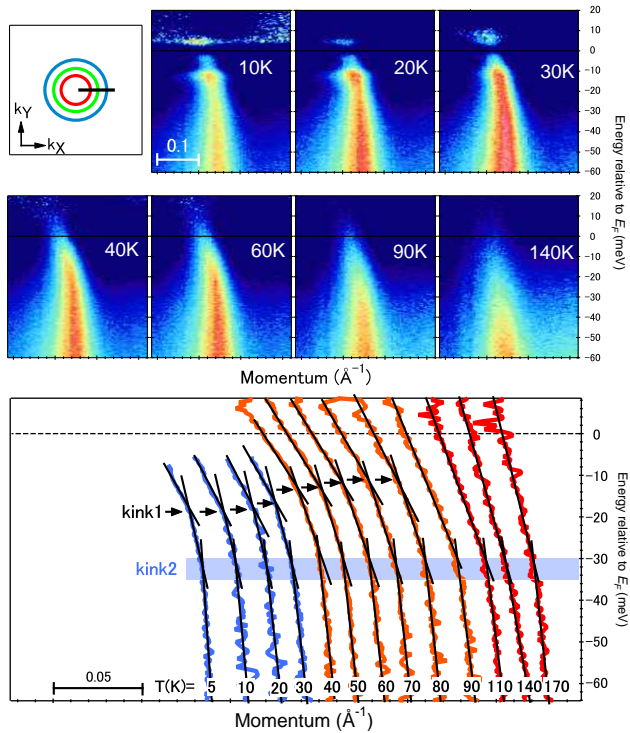}
\caption{\label{fig3} (color online) Temperature-dependent Laser ARPES data of OP BaK with $x \sim$ 0.4: (upper panel) $E-k$ plots of a cut taken near the high-symmetry line with v-polarized light at different temperatures. The data are divided by FD function. $k_X$, $k_Y$ axes are along the Fe-As bond. (lower panel) Temperature dependence of the kink energy in BaK with $x \sim$ 0.4. The colored curves represent the dispersion corresponding to MDC peak positions determined from Lorentzian fitting at different temperatures. The kink energy position was estimated from the intersection points between two different slopes in the dispersion as in Fig. \ref{fig1}: The low-BE kink1 energy position is marked by a black arrow and the high-BE kink2 is marked by a thick blue horizontal line.}
\end{center}
\end{figure}



Now regarding the difference in the temperature-dependence of kink1 in our ARPES results from one side and the mode deduced from INS \cite{Castellani_PRL11} and STM  \cite{HHWen_NatPhys13} data from the other side, it does not necessarily imply differences in the origin of the same mode which could be observed by these three experimental techniques. It may be just a technical difference because in most of INS and STM studies \cite{Castellani_PRL11, HHWen_NatPhys13}, the high-temperature data were taken only slightly above $T_c$ with no clear evidence of the complete disappearance of this mode at temperatures high enough above $T_c$. However, in our ARPES study we have extended the measurements to temperatures much higher than $T_c$ ($\sim$ 170K) and confirmed the survival of the mode above $T_c$, namely up to $\sim$ 90K. The doping-dependence agreement that we presented is much stronger than this temperature-dependence disagreement and it will not change our conclusion of a magnetic origin of this mode. Indeed, spin fluctuations (excitations) can survive above $T_c$ although weaker than the SC state as pointed out by some studies.




In summary, our Laser ARPES results on a wide doping range of BaK compound uncovered more details about the spectral line shape and energy scales in this compound: a robust low-BE kink structure in the dispersions which is doping dependent with its energy peaking at the OP level ($x \sim$ 0.4) and decreasing towards the UD and OD sides. It is also temperature-dependent and survives at temperatures higher than $T_c$. We attribute this kink structure to electron-mode coupling in good agreement with INS and STM data on the same compound which observed a bosonic mode at a similar energy scale and was considered to be related with spin excitations. The relation between the mode energy ($\Omega$) and the SC transition temperature ($T_c$) deduced from our BaK Laser ARPES data follows the universal relation deduced from INS and STM. We could also resolve an additional high-BE kink2 which is less doping- and temperature-dependent compared to kink1 and may thus be of different origin.



\section*{Acknowledgements}
We thank Y.~Matsuda, T.~Shibauchi, K.~Kuroki, K.~Kontani, R.~Arita, S.~Onari and S.~Shamoto for informative discussions. This research
is supported by the Japan Society for the Promotion of Science
(JSPS) through its Funding Program for World-Leading Innovation R
and D on Science and Technology (FIRST) program. This work was partly supported by Strategic International Collaborative Research Program (SICORP), Japan Science and Technology Agency (JST).




\end{document}